\documentclass[journal]{IEEEtran}
\usepackage[utf8]{inputenc}
\usepackage{amsmath}
\usepackage{amsfonts}
\usepackage{amssymb}
\usepackage{graphicx}
\usepackage{amsmath}
\usepackage{enumitem}   
\usepackage{cite}
\usepackage{comment}
\newcommand{\argminD}{\arg\!\min} 
\usepackage{comment}
\usepackage{subcaption}
\graphicspath{ {Figures/} }
\title{Target detection using underwater acoustic communication links}
\author{
\IEEEauthorblockN{Lu Shen\IEEEauthorrefmark{1}, Yuriy Zakharov\IEEEauthorrefmark{1}, Benjamin Henson\IEEEauthorrefmark{1}, Nils Morozs\IEEEauthorrefmark{1}, \\Beno\^it Parrein\IEEEauthorrefmark{2}, Paul D. Mitchell\IEEEauthorrefmark{1}}\\
 \IEEEauthorblockA{\IEEEauthorrefmark{1}School of Physics, Engineering and Technology, University of York, U.K}
    \\
    \IEEEauthorblockA{\IEEEauthorrefmark{2}Nantes Universit\'e, Polytech Nantes, CNRS, LS2N, F-44000 Nantes, France}
}
\begin{document}
\maketitle

\begin{abstract}
Underwater monitoring and surveillance systems are essential for underwater target detection, localization and classification. 
The aim of this work is to investigate the possibility of target detection by using data transmission between communication nodes in an underwater acoustic (UWA) network, i.e, re-using acoustic communication signals for target detection.
A new target detection method based on estimation of the time-varying channel impulse response between the communication transmitter(s) and receiver(s) is proposed and investigated. This is based on a lake experiment and numerical experiments using a simulator developed for modeling the time-varying UWA channel in the presence of a moving target. 
The proposed detection method provides a clear indication of a target crossing the communication link.
A good similarity between results obtained in the numerical and lake experiments is observed.
\end{abstract}


\section{Introduction}
%
Underwater monitoring and surveillance systems are essential for underwater target detection, localization and classification. Dedicated passive and active sonar systems are commonly used for this purpose. The detection performance of passive systems is limited when the acoustic sound produced by the target is of a low intensity, in which case active sonars, monostatic or bistatic, are preferable. Monostatic sonars are based on backscattering of acoustic signals from the target. It is however known that the forward scattering from the target, exploited in bistatic sonars, may have a higher intensity~\cite{urick1983principles}.

Active multiple-input multiple-output (MIMO) sonar systems can significantly improve target detection performance~\cite{pailhas2016spatially, pailhas2017tracking}.  When the number of MIMO links is large, it is possible to detect the target with a single MIMO snapshot~\cite{pailhas2016spatially}. 

A number of lake experiments were conducted between 1997 and 1999 to investigate methods of detecting forward scattering from a moving target~\cite{zverev2001experimental, matveev2002determination}. In these experiments, tone signals are simultaneously transmitted at a set of carrier frequencies within the frequency range 1--3~kHz. A signal source fixed at the lake bottom or a vertical antenna array is used for transmission. Both vertical and horizontal receiving arrays are deployed at a distance of about 400~m. Targets of several meters in length and up to 1~m height are towed underwater across the area between the transmitter and receiving arrays. The aim is to detect the diffracted signals against the fluctuating direct signal.
In~\cite{zverev2001experimental}, 
fluctuations of the direct signal are removed by a high-pass filter, the cut-off frequency of which is computed based on \textit{a priori} information, including the size and velocity of target. 
The signal then goes through a filter matched to a theoretical model of the diffracted signal and incoherently accumulated over the array elements and the carrier frequencies. The target detection is performed based on the variation of the output signal. The performance is evaluated by the output signal-to-noise ratio ($\mathrm{SNR}_\mathrm{out}$) which is defined as the ratio between an increment in the output signal and the standard deviation of the output signal without a target.
It is found in~\cite{matveev2002determination} that $\mathrm{SNR}_\mathrm{out}$ can be improved by jointly processing the output signals of two vertical arrays. Another finding is that it is possible to indicate the direction of the target movement based on the shape of the peaks in the processed signal.
A comparative analysis of detection methods is conducted in~\cite{matveev2005comparative} using the experimental data described in~\cite{zverev2001experimental, matveev2002determination}. 
It is found that the use of a vertical array shows more robust target detection performance compared to a horizontal array. 
Further analysis is conducted in~\cite{matveev2007forward} to investigate the detection performance with coherent, partially coherent and incoherent spatial processing. It is indicated that the choice of spatial processing method depends on the vertical distribution of the acoustic field.
Aforementioned works focus on investigating matched filtering based methods for detecting the forward scattering from a target using transmit and/or receive antenna array and low-frequency pulse transmission.

In~\cite{lei2017detection}, an adaptive cancellation algorithm is applied to the received signals to enhance the detection of forward scattering from the strong direct path in a lake experiment. The source is located at 4~m depth. A vertical array with 10 hydrophones is deployed at a distance of 900~m.
The dimension of the target is 6~m$\times$1~m.
Successive pulses at a 10~kHz carrier frequency are transmitted. 
An adaptive filter is used for signal processing.
The input of the adaptive filter is a segment of a signal received when the target is absent. Other segments of the received signal are used as the desired signal, with which the filter output is compared to produce the adaptive filter error. 
As the forward scattered signal is weakly correlated with the direct signal, it is assumed that the algorithm will only cancel the direct signal, and the adaptive filter error will be higher when the target is present. This method also requires a vertical antenna array.

In this work, we consider scenarios with a distributed network of multiple unsynchronized acoustic sensor nodes capable of communicating with each other, instead of using source/receiver arrays. 
Such a network provides flexibility in the deployment of the acoustic sensor nodes based on the environment.
Wideband acoustic signals at a relatively high frequency (32~kHz carrier frequency in our experiments) are used for data transmission. 
The aim of this work is to investigate the possibility of re-using the communication signals in the network for target detection. 
The target detection is performed based on the measurement of the time-varying channel impulse responses (CIRs) between communication nodes in the network. Note that the channel estimation, i.e. CIR measurement, is one of the main steps of signal processing in the receiver, hence an CIR estimate is normally available at the receiver. The idea is to detect changes in the CIR caused by the target movement. The use of wideband signal provides high resolution CIRs, which increases the chance of target detection.
In this paper, we focus on the target detection performed for each individual  network link. Designs of multiple access control and network layer protocols for such a target detection network are presented in~\cite{morozs2023target, morozs2023network}.

\begin{figure}
\centering
\includegraphics[width=\linewidth]{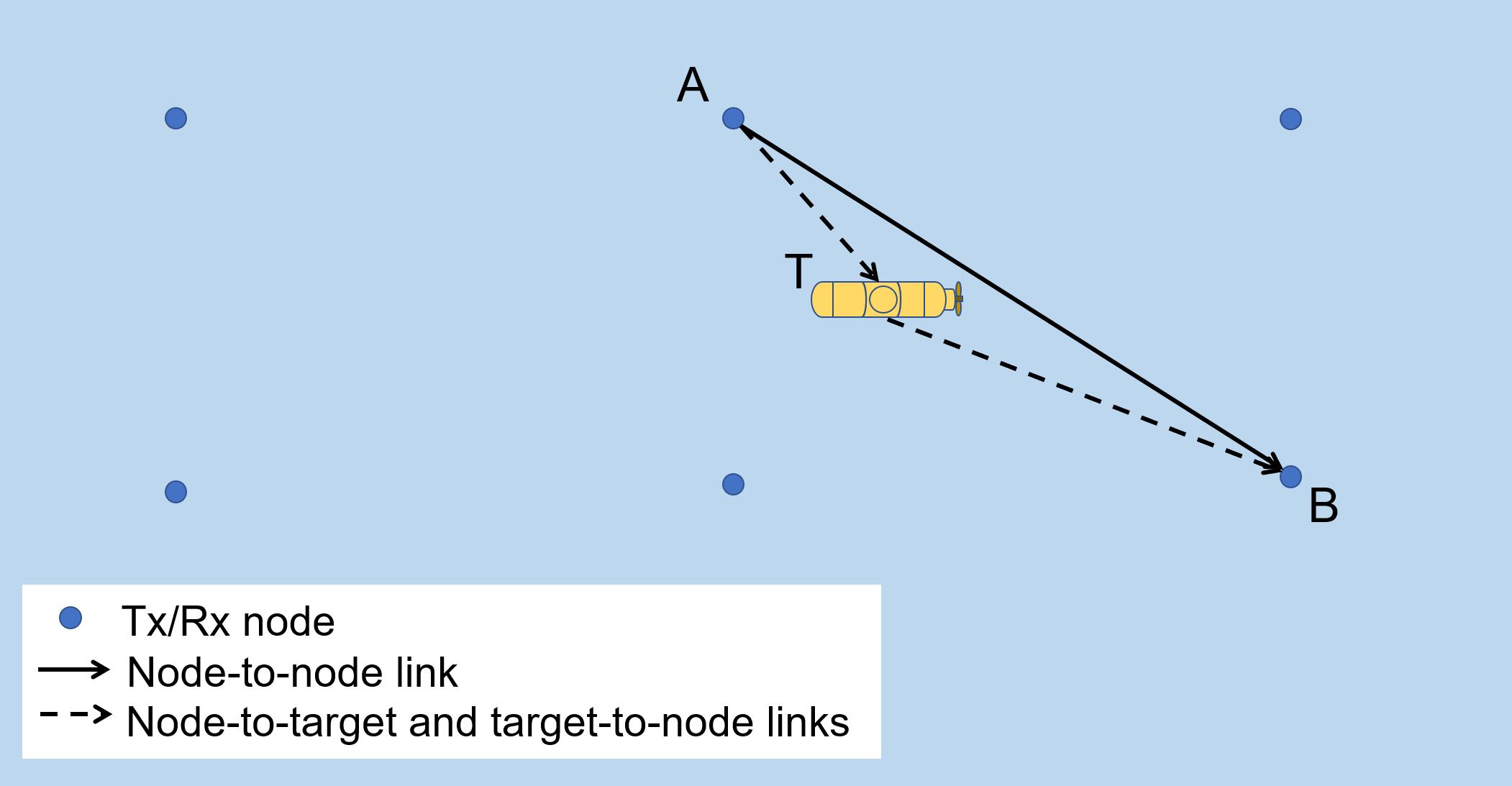}
\caption{Acoustic propagation between two nodes (A and B) with presence of a target (T) in a distributed network (Top view).\label{Fig:scene}}
\end{figure}

Fig.~\ref{Fig:scene} illustrates a scene of a distributed network with six Tx/Rx nodes and a target. We consider the scenario when there is a `silent' target moving underwater. When the target is absent, the acoustic multipath propagation from node A to node B only includes the node-to-node link. When there is a target, apart from the link between two nodes, scattering from the target will also be present. 
The acoustic propagation between two nodes in the presence of the target will correspond to a different CIR compared to the `no target' case. Therefore, it is possible to perform target detection based on changes in the CIR in time. A challenge is to distinguish between changes in the CIR due to the environmental factor, such as node drift or surface waves, and change caused by a moving target. 
To this end, a detection algorithm is proposed in this paper which reduces the impact of the environmental variation on the CIR and subsequent detection performance.

For performance evaluation of the target detection scheme, an acoustic simulator for conducting virtual signal transmission experiments is developed; it is capable of modelling the multipath propagation and Doppler effects for different UWA environments taking into account the motion of the transmitter/receiver/target. One of the key requirements of the simulator is to generate highly accurate acoustic field information for the communication links based on the trajectories of the nodes and the target. Unlike the typical static/slowly moving transmitter and receiver nodes, the target might be moving at a high speed. To track the fast variation of the acoustic channel, a high sampling rate of channel is required. Therefore, the simulator computes the acoustic field at the baseband sampling interval of the communication signals. To reduce the computational time, the simulator exploits the principles of the Waymark and VirTEX simulators~\cite{liao2017grid, siderius2008modeling}, where the acoustic field is pre-computed within an area of interest on a grid of space points and then interpolated based on the pre-computed information.

The field interpolation method proposed in~\cite{siderius2008modeling} assumes that the receiver is in the far field with respect to the signal source. However, the target can reflect/re-radiate a transmission and could be positioned closely to the communication nodes, thus the interpolation method used in~\cite{liao2017grid, siderius2008modeling} might not be accurate enough.
To improve the field interpolation in the near field, an improved interpolation approach is proposed. 
The proposed simulator is used to generate the time-varying CIR and the signal at the receiver. 


To verify the target detection performance and validate the simulator, a lake experiment is conducted with several nodes and a target whose trajectory is accurately estimated using acoustic navigation.  In the lake experiments, the CIR estimates are obtained from the channel estimator in the receiver and then used for the target detection.
After the lake experiment, we run the simulator with the same node configuration, and use the target trajectory estimate as input to generate the CIRs and received signals. A good match between the experimental and simulation results is observed.

The contributions of this paper are as follows.
\begin{itemize}
\item A target detection algorithm is proposed based on measuring changes in the time-varying CIRs between communication nodes, which are estimated using the existing communication signals in the network.
\item An acoustic simulator is developed, which takes into account the motion of the target and the communication nodes. The output of the simulator includes the signals received at nodes and time-varying CIRs between each transmitter-receiver pair.
An approach is proposed and implemented to improve the acoustic field interpolation in the near field. The simulator is validated using the lake experiment.
\item The performance of the proposed detection method is investigated using numerical and lake experiments.
\end{itemize}

The structure of the paper is as follows. Section~\ref{sec:target_detection} describes the target detection algorithm. Section~\ref{sec:general_structure} introduces the acoustic simulator. Section~\ref{sec:interpolation} describes the acoustic field interpolation approaches. Numerical simulation and lake experiments are presented in Section~\ref{sec:numerical_simulation} and \ref{Sec:lake_exp}, respectively. The paper is concluded in Section~\ref{sec:conclude}.

\section{Target detection based on CIR variation}\label{sec:target_detection}
The key idea of target detection is to identify changes in the CIR caused by a moving target. 
The changes in the CIR can be measured by the mean squared deviation (MSD)~\cite{sayed2003fundamentals}:
\begin{equation}
\mathrm{MSD} = \sum_{n = 0}^{L-1}|h_0(n) - h(n)|^2,\label{eq:MSD}
\end{equation}
where $L$ is the length of the CIR, $h_0(n)$ is the reference CIR when the target is absent, $h(n)$ is the CIR of interest, obtained at a time interval where a target might be present.
The MSD in (\ref{eq:MSD}) can be computed using the frequency-domain representation of the channel:
\begin{equation}
\mathrm{MSD} = \sum_{k = 0}^{K-1}|H_0(k) - H(k)|^2,
\end{equation}
where $H_0, H$ are the frequency responses of filters $h_0$ and $h$, respectively, and they are computed using the fast Fourier Transform (FFT) of a size $K$. When the MSD is a small value, we can assume that the target is absent. For high MSD values, we can assume that the target is present.

However, even when the target is absent, the CIR will be time-varying due to the time-varying nature of the UWA channel, thus resulting in a high MSD. Therefore, we need to distinguish between the change of the impulse response due to the time-varying environment and the change caused by a moving target. Instead of using $h_0$ in (\ref{eq:MSD}), we can use $h_0*g$, where $*$ denotes the convolution operation and $g$ is an impulse response representing a filter with a small delay spread. In such a case, the change in delay and amplitude caused by the slow motion of the source and the receiver, the variation of the environment and the clock mismatch between the transmitter and the receiver can be eliminated from influencing the MSD.
When a target is crossing the communication link, the multipath structure of the propagation channel will change significantly, and these changes cannot be compensated by the filter $g$.

Small delays introduced by the filter $g$ in the time domain correspond to complex exponentials of long periods in the frequency domain.
Therefore, the frequency response of $g$ can be expressed as a combination of complex exponentials with expansion coefficients $c(p)$~\cite{yu2015iterative}:
\begin{equation}
G(k) = \sum_{p = 0}^{2P} c(p)e^{j2\pi (-P+p)k/K}, \,\, k = 0, \ldots, K-1, \label{eq:G_k}
\end{equation}
where the number of complex exponentials $2P+1$ corresponds to the delay spread of the filter $g$.

Therefore, to find the filter $g$, we solve the following optimization problem:
\begin{equation}
\hat{G} = \argminD_{G}\sum_{k = 0}^{K-1}|H_0(k)G(k) - H(k)|^2.\label{eq:op}
\end{equation}
We denote the combined frequency response as:
\begin{equation}
\begin{split}
\tilde{H}(k) &= H_0(k)G(k)  \\
             &= \sum_{p = 0}^{2P}c(p)\tilde{G}(k, p),\label{eq:H_k}
\end{split}
\end{equation}
where $\tilde{G}(k, p) = H_0(k)e^{j2\pi(-P+p)k/K}$, and find the expansion coefficients $\mathbf{c} = [c(0), \ldots, c(2P)]^T$ by minimizing the cost function:
\begin{equation}
J(\mathbf{c}) = \sum_{k = 0}^{K-1}|\tilde{H}(k) - H(k)|^2.
\end{equation}
This is achieved by solving the normal equation:
\begin{equation}
(\mathbf{R} + \varepsilon\mathbf{I})\mathbf{c} = \mathbf{b},\label{eq:normal_eq}
\end{equation}
where $\mathbf{I}$ is an identity matrix, $\mathbf{R} = \mathbf{\tilde{G}}^H \mathbf{\tilde{G}}$, $\mathbf{b} = \mathbf{\tilde{G}}^H \mathbf{H}$, $\mathbf{\tilde{G}}$ is a $K\times (2P+1)$ matrix with elements $\tilde{G}(k, p)$, $\mathbf{H} = [H(0), \ldots, H(K-1)]^T$ and $\varepsilon > 0$ is a regularization parameter.
The solution of (\ref{eq:normal_eq}) is given by:
\begin{equation}
\hat{\mathbf{c}} = (\mathbf{R} + \varepsilon\mathbf{I})^{-1} \mathbf{b}.
\end{equation}
After obtaining $\hat{\mathbf{c}}$, we have $\tilde{H}(k) = \sum_{p = 0}^{2P}\hat{c}(p)\tilde{G}(k, p)$. The deviation in the frequency response is computed as: $\Delta H(k) = \tilde{H}(k) - H(k)$.
Finally, the normalized MSD is computed by: 
\begin{equation}
\mathrm{MSD}_\mathrm{norm} = \dfrac{\sum_{k = 0}^{K}|\Delta H(k)|^2}{\sum_{k = 0}^{K}|H_0(k)|^2}.\label{eq:norm_MSD}
\end{equation}

The normalized MSD is used as the statistic for the target detection. When it is small, it means that the communication link is not distorted, i.e, the target is absent. When it is high, it can be inferred that the link distortion is caused by a target.

\section{Simulator for acoustic links}\label{sec:general_structure}

Underwater acoustic channel simulators, such as Waymark~\cite{ liao2017grid,liu2012doubly, henson2014waymark}, can model the signal transmission with a moving transmitter and receiver.
To reduce the computational complexity, the acoustic field is sampled at an interval, which is typically significantly longer than the baseband sampling interval~\cite{liao2017grid, henson2014waymark}. In this work, we introduce a moving target crossing the communication links, and to guarantee accurate modelling of the signal transmission in the presence of a fast-moving target, the acoustic field is computed at the baseband sampling rate. 
Even at baseband frequencies, generating the acoustic field for prolonged scenarios can be impractical for meaningful numerical experiments.
Therefore, the simulator exploits the principles used in~\cite{liao2017grid, siderius2008modeling} and performs the task in two stages: pre-processing and received signal computation. 

\begin{figure*}
\centering

\begin{subfigure}{0.49\textwidth}
\vspace{10pt}
    \includegraphics[width=\textwidth]{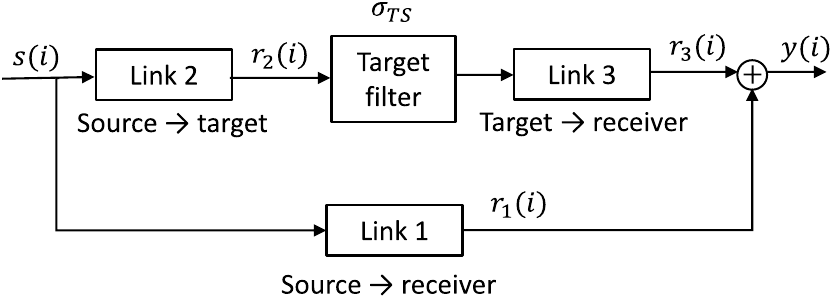}
    \caption{}
    \label{Fig:filtering}
\end{subfigure}
\hfill
\begin{subfigure}{0.49\textwidth}
    \includegraphics[width=\textwidth]{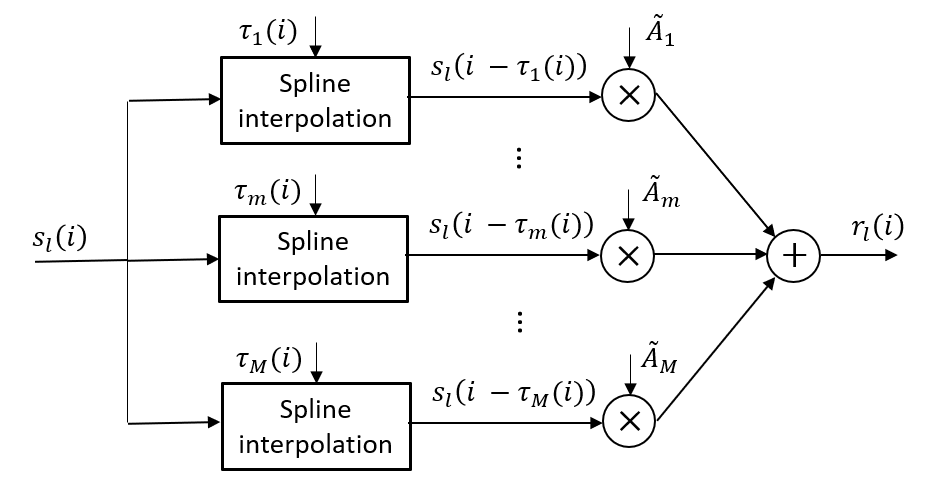}
    \caption{}
    \label{Fig:filtering_link}
\end{subfigure}

\caption{General structure of the simulator: (a) Received signal computation; (b) Filtering for the $l$th link, $l = 1, 2, 3$, $i$ is the index of the baseband sample, where $s_l$ is the baseband input signal and $r_l$ is the baseband output. }
\label{Fig:simulator_structure}
\end{figure*}

\subsection{Pre-processing}\label{sec:pre-processing}

The acoustic field between a pre-defined source position and possible receiver positions within an area of interest on a predefined set of grid points is pre-computed. The BELLHOP3D ray-tracing program~\cite{porter2016bellhop} is used to generate the acoustic field information, including the delay $\tau$, amplitude $A$, departure angle $\theta_d$ and incident angle $\theta_i$ of each arrival for each grid point. We consider a simplified environment with a flat surface, flat bottom and range-invariant sound speed profile.
The pre-computation is performed for a set of source depths required to cover the range of possible source depths.
For each source depth, acoustic field computation is performed on a two-dimentional (depth, range) grid map of receiver positions. For the source parameters, the anticipated source positions and target trajectories define the minimum depth, the maximum depth and the resolution (step size) in depth $\Delta D_s$. The parameters of the grid map include the minimum and maximum range, the minimum and maximum depth, and the resolution of the grid map in range and in depth. 

\subsection{Received signal computation}\label{sec:online-processing}
To obtain the received signal, the acoustic propagation of three links need to be considered as illustrated in Fig.~\ref{Fig:filtering}. Link~1 is the direct link between the source and the receiver when the target is absent; Link~2 is the link between the source and target; and Link~3 is the link between the target and receiver. The target plays the role of a receiver when modelling link~2, and it acts as a source when modelling link~3. A target filter is used to incorporate a target strength model~\cite{goddard2008sonar}. 

The acoustic field of all three links are computed by acoustic field interpolation.   
The trajectories of the source, receiver and target are interpolated to the baseband sampling rate. 
The field interpolation is performed according to the source, target and receiver positions with respect to the closest grid points within the grid maps using the field information at these grid points. More details on the acoustic field interpolation are given in Section~\ref{sec:interpolation}.

As shown in Fig.~\ref{Fig:filtering_link}, the output of the $l$th link is the signal:
\begin{equation}
r_l(i) = \sum_{m = 1}^{M}\tilde{A}_m(i)s_l(i - \tau_m(i)),
\end{equation}
where $i$ is the index of a baseband sample, $s_l(i)$ is the baseband input of the $l$th link, $M$ is the number of arrivals, $\tau_m(i)$ is the delay of the $m$th arrival obtained from the field interpolation, and $\tilde{A}_m(i)$ is the baseband equivalent amplitude of the $m$th arrival computed according to beampatterns of the source and receiver:
\begin{equation}
\tilde{A}_m(i) = G_s(\theta_d^m)G_r(\theta_i^m)A_m(i)e^{-j\omega_c\tau_m(i)},
\end{equation}
where $\omega_c = 2\pi f_c$, $f_c$ is the carrier frequency of the transmitted signal, $A_m(i)$ is the amplitude of the $m$th arrival obtained from the field interpolation, $G_s$ and $G_r$ are the source and receiver beampatterns, $\theta_d^m$ is the departure angle of the $m$th arrival, $\theta_i^m$ is the incident angle of the $m$th arrival, obtained from the field interpolation. 
The baseband input sample $s_l(i - \tau_m(i))$ is obtained using the local spline interpolation as described in~\cite{liu2012doubly}.
The signal at the receiver is then given by:
\begin{equation}
y(i) = r_1(i) + r_3(i).
\end{equation}

In this paper, the target is modelled as a rigid sphere of radius $a$. The bistatic target strength can be written as a function of the angle $\alpha (i)$ between the incident and scattered rays. Here we compute $\alpha (i)$ as:
\begin{equation}
\alpha (i) = |\alpha_\mathrm{ST}(i) - \alpha_\mathrm{TR}(i)|,
\end{equation}
where $\alpha_\mathrm{ST}(i)$ is the azimuth angle between the source and the target and $\alpha_\mathrm{TR}(i)$ is the azimuth angle between the target and receiver and $ \alpha_\mathrm{TR}(i) \in (-\pi, \pi]$.
Therefore, the target filter is modelled as a single-tap filter with a target strength coefficient~\cite{sonar_handbook}:
\small
\begin{equation}
\sigma_{TS}(i) = 
\begin{cases}
\dfrac{a^2}{4}\left[ 1+\tan^2\left(\dfrac{\alpha (i)}{2}\right)J_1^2\left(wa\sin \alpha (i)\right) \right] & \alpha (i) \neq \pi \\
\dfrac{a^2}{4}(1+w^2 a^2) & \alpha (i) = \pi 
\end{cases} \label{eq:target_sphere}
\end{equation}
\normalsize
where $w = 2\pi/\lambda$ is the wavenumber, $\lambda$ is the acoustic wavelength, $J_1(\cdot)$ is the Bessel function of the first order.

\subsection{Channel impulse response computation}\label{sec:channel_impulse_response}
Apart from the signal at the receiver, the simulator also provides the time-varying CIR between the source and receiver. 
For every position of the signal source, receiver and target, the channel frequency response is computed as:
\begin{equation}
H(i, k) = H_1(i, k) + \sigma_{TS}(i)H_2(i, k)H_3(i, k),\label{eq:channel_freq_response}
\end{equation}
where $H_1$, $H_2$, $H_3$ represent the baseband frequency responses of the three links in Fig.~\ref{Fig:filtering}. For instance, the frequency response $H_1(i,k)$ is computed as:
\begin{equation}
H_1(i,k) = \sum_{m = 1}^{M}\tilde{A}_m(i)e^{-j2\pi k\Delta f \tau_m(i)}, \,\,\, k \in [-B/2, B/ 2], \label{eq:H1}
\end{equation}
where $B\Delta f$ is the frequency bandwidth, $\Delta f$ is the frequency resolution. The channel impulse response $\mathbf{h}(i)$ is then computed using the inverse FFT of $H$.

\section{Acoustic field interpolation}\label{sec:interpolation}
For pre-computation, a set of grid maps are generated, each corresponding to a particular source depth. 
When the source/receiver is not on a grid point, acoustic field interpolation is performed.
Plane wave acoustic field approximation works well when the receiver is in the far field of the signal source. 
In simulation scenarios, the target (as a receiver and signal source) can be arbitrarliy close to a node; in such cases, the spherical wave approximation improves the interpolation~\cite{li2023data}.
In this section, we present acoustic field interpolation based on plane wave and spherical wave approximations.

\subsection{Source at a depth with precomputed grid map and receiver between grid points}\label{subsec:receiver_deviate}
We first consider a scenario when the source is at a depth for which a grid map has been precomputed and the receiver is located at a point $(x, y)$ between grid points on the map. 
\subsubsection{Plane wave interpolation}
\begin{figure}
\centering
\includegraphics[width=0.8\linewidth]{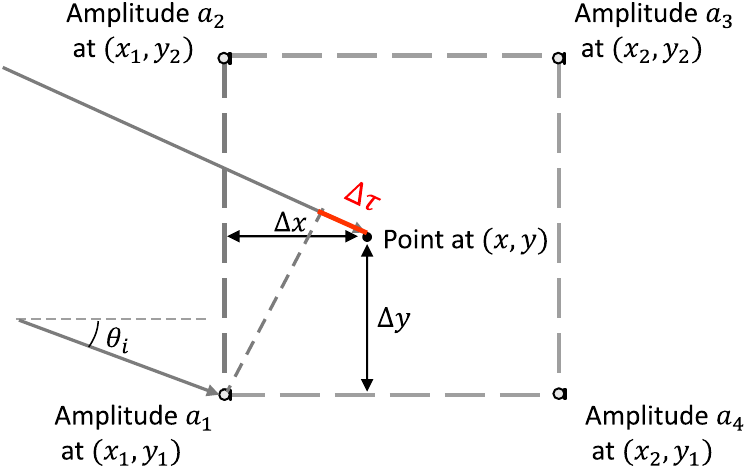}
\caption{An illustration of plane wave interpolation for a receiver position $(x, y)$ based on the ray information at the grid point $(x_1, y_1)$, where $\theta_i$ is the incident angle of a ray arriving at $(x_1, y_1)$. \label{Fig:rx_plane_interp}}
\end{figure}
\begin{figure*}
\centering
\includegraphics[width=0.77\linewidth]{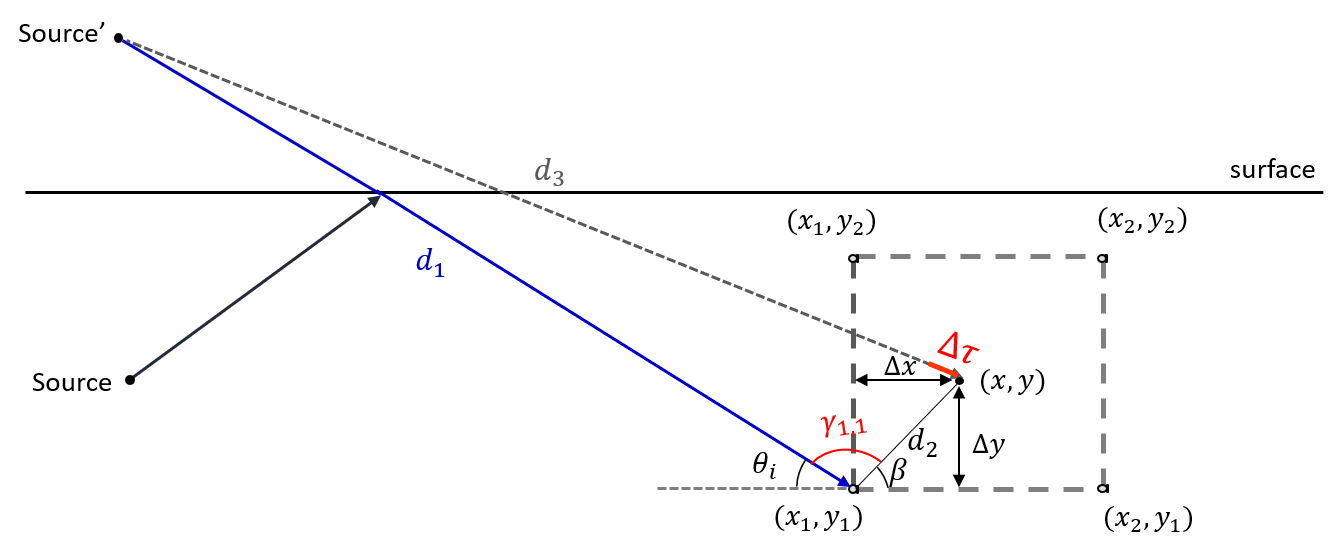}
\caption{An illustration of spherical wave interpolation for a receiver position $(x, y)$ based on the ray information at the grid point $(x_1, y_1)$, where $\theta_i$ is the incident angle of a ray arriving at $(x_1, y_1)$, $\gamma_{1, 1}$ is the angle between $d_1$ and $d_2$, and $\beta$ is the angle bewteen $d_2$ and $x$ positive axis.\label{Fig:rx_sphere_interp}}
\end{figure*}
In~\cite{siderius2008modeling}, a plane wave assumption is used for acoustic field interpolation as shown in Fig.~\ref{Fig:rx_plane_interp}. 
The arrivals at the receiver position $(x, y)$ is the combination of the arrivals at the four neighbouring grid points $(x_a, y_b), a,b \in [1,2]$, with weighted amplitudes and delay adjustments. 
The delays are adjusted by the difference in the wave travel time between the neighbouring points and the position of the receiver.
Assuming that a ray is arriving at a grid point $(x_a, y_b)$ with an incident angle $\theta_i$, the difference in the travel time is given by~\cite{siderius2008modeling}:
\begin{equation}
\Delta \tau = (\Delta x \cos{(\theta_i)} + \Delta y \sin{(\theta_i)})/c, \label{eq:plane_rx_delay}
\end{equation}
where $c$ is the speed of sound, $\Delta x = x - x_a$ and $\Delta y = y - y_b$. 
The delay of arrival for the $m$th ray from $(x_a, y_b)$ is updated as: $\tau_m^{'} = \tau_m + \Delta \tau$.


The amplitude of the arrivals for the four neighbouring grid points are weighted by:
\begin{equation}
\begin{aligned}
&  a_1:\, (1-w_1)(1- w_2), \\
&  a_2:\, (1-w_1)w_2, \\
&  a_3:\, w_1 w_2, \\
&  a_4:\, w_1 (1- w_2), 
\end{aligned} \label{eq:coeff_amp}
\end{equation}
where $a_1$, $a_2$, $a_3$, $a_4$ represent amplitudes of the arrivals at the four grid points. The weights are computed based on the position of the receiver $(x, y)$ with respect to the grid points:
\begin{equation}
\begin{aligned}
w_1 &= (x - x_a)/(x_2 - x_1),\\
w_2 &= (y - y_b)/(y_2 - y_1).
\end{aligned} \label{eq:weight}
\end{equation}

The main steps of the plain wave interpolation can be summarized as:
\begin{enumerate}[label=(\roman*)]
\item Adjust the delay of the arrivals at four grid points based on  (\ref{eq:plane_rx_delay}).
\item Apply weighting to the amplitudes of the arrivals at four grid points according to (\ref{eq:coeff_amp}) and (\ref{eq:weight}).
\end{enumerate}

\begin{figure*}
\centering
\begin{subfigure}{0.35\textwidth}
\vspace{20pt}
\centering
\includegraphics[width=\textwidth]{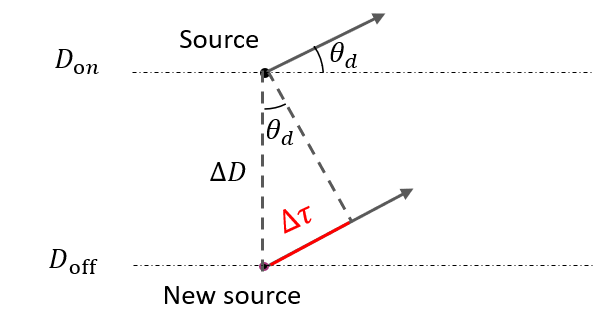}
\caption{\label{Fig:tx_plane_interp}}
\end{subfigure}
\hfill
\begin{subfigure}{0.53\textwidth}
\includegraphics[width=\textwidth]{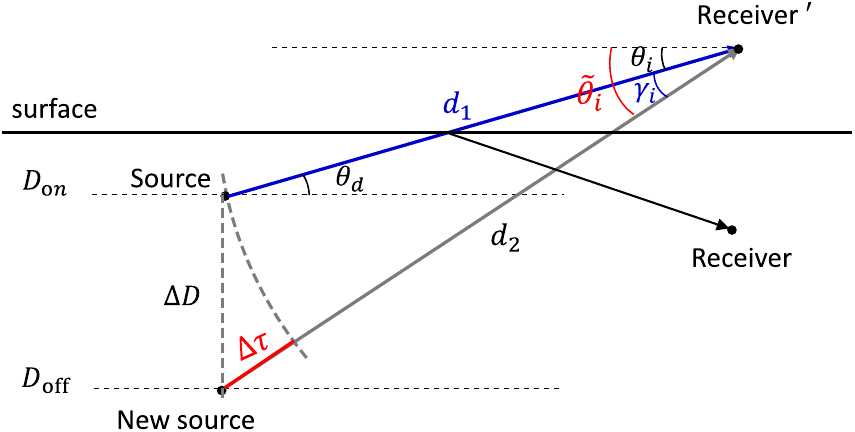}
\caption{\label{Fig:tx_sphere_interp}}
\end{subfigure}
\caption{Illustrations of interpolation schemes for new source depth based on the ray information at the known source depth; (a) plane wave interpolation, $\theta_d$ is the departure angle of a ray travelling from the original source; (b) spherical wave interpolation, $\tilde{\theta}_i$ is an incident angle due to the change in source depth.}
\end{figure*}

\subsubsection{Spherical wave interpolation}
An illustration of delay adjustment when using spherical wave approximation is shown in Fig.~\ref{Fig:rx_sphere_interp}. As the delay adjustment for a direct path is straightforward, we use a surface reflection as an example. The same procedure applies for the bottom reflections.

We first find the image source position ($\mathrm{Source}^{'}$) using the image source method (ISM)~\cite{allen1979image}. As shown in Fig.~\ref{Fig:rx_sphere_interp}, $d_1$ is the distance a ray travels to arrive at $(x_1, y_1)$ and $d_3$ is the distance it needs to travel to arrive at $(x, y)$. The delay adjustment is made based on the difference between $d_1$ and $d_3$.

For the $m$th arrival at a grid point $(x_a, y_b)$, the distance between the image source position and the grid point can be computed as:  $d_1 = c\tau_m$, where $\tau_m$ is the delay of the $m$th arrival. The distance between the receiver position $(x, y)$ and $(x_a, y_b)$ can be found as: $d_2 = \sqrt{(\Delta x)^2 + (\Delta y)^2}$. By denoting the angle  between $d_1$ and $d_2$ as $\gamma_{a, b}$, and assuming the angle to be known, distance $d_3$ can be computed based on the cosine rule as:
\begin{equation}
d_3 = \sqrt{d_1^2 + d_2^2 - 2d_1 d_2 \cos{(\gamma_{a,b})}}.\label{eq:d3}
\end{equation} 
Thus, we can obtain the change in the delay $\tau_m$ as:
\begin{equation}
\Delta \tau  = (d_3 - d_1)/c, \label{eq:spherical_delay_adjust}
\end{equation}
and the delay of the $m$th arrival at $(x, y)$ can be updated as: $\tau_m^{'} = \tau_m + \Delta \tau$.

The angle $\gamma_{a,b}$ can be computed using the incident angle of the arrival $\theta_i$ and the angle between $d_2$ and $x$ positive axis (denoted as $\beta$). 
Note that the incident angle $\theta_i$ is within $[-\pi/2, \pi/2]$.
For a grid point $(x_a, y_b)$:
\begin{equation}
\beta = \arctan{\left(\dfrac{|y - y_b|}{|x - x_a|}\right)}.
\end{equation}
For the four grid points, we have:
\begin{equation}
\begin{aligned}
\gamma_{1, 1} &= 
\begin{cases}
\pi - \mathrm{sign}(\theta_i + \beta)(\theta_i + \beta) & \theta_i < 0 \\
\pi - (\theta_i + \beta) & \theta_i \geq 0
\end{cases}
\\
\gamma_{1, 2} &= 
\begin{cases}
\pi - \mathrm{sign}(\theta_i - \beta)(\theta_i - \beta) & \theta_i > 0 \\
\pi + \theta_i - \beta & \theta_i \leq 0
\end{cases}
\\
\gamma_{2, 1} &= 
\begin{cases}
\mathrm{sign}(\theta_i - \beta)(\theta_i - \beta) &\qquad \theta_i > 0 \\
\theta_i - \beta & \qquad\theta_i \leq 0
\end{cases}
\\
\gamma_{2, 2} &= 
\begin{cases}
\mathrm{sign}(\theta_i + \beta)(\theta_i + \beta) &\qquad\theta_i < 0 \\
\theta_i + \beta &\qquad \theta_i \geq 0
\end{cases}
\end{aligned}
\label{eq:gamma}
\end{equation}
The amplitudes of the arrivals are weighted according to (\ref{eq:coeff_amp}) and (\ref{eq:weight}).

\subsection{Source at a depth without precomputed grid map and receiver at a grid point}\label{subsec:source_deviate}

Now we consider the scenario when a source depth deviates from the source depth, for which a grid map exists, and the receiver position is on a grid point of the map.  
The idea is to adjust the delay of each arrival based on the deviation of the source depth from the nearest source depth with a precomputed grid map. 

\subsubsection{Plane wave interpolation}
As illustrated in Fig.~\ref{Fig:tx_plane_interp}, the difference in the travel time can be computed as:
\begin{equation}
\Delta \tau = -\Delta D \sin{(\theta_d)}/c, \label{eq:plane_src_delay}
\end{equation}
where $\Delta D = D_\mathrm{off} - D_\mathrm{on}$, $D_\mathrm{off}$ is a source depth not with a pre-computed grid map and $D_\mathrm{on}$ is the nearest source depth with a grid map.

\subsubsection{Spherical wave interpolation}
We first find the image receiver position based on ISM as shown in Fig.~\ref{Fig:tx_sphere_interp}.  Assuming that there is a ray from the depth $D_\mathrm{on}$ with a departure angle of $\theta_d$ arriving at the receiver with a delay $\tau_m$, the distance $d_1$ between the source and the image receiver can be computed as: $d_1 = \tau_m c$.  Based on the cosine rule,  we have:
\begin{equation}
d_2 = 
\begin{cases}
\sqrt{d_1^2 + (\Delta D)^2 -2d_1 |\Delta D| \cos{(\pi/2 - \theta_i)}}& \theta_d > 0, \\
\sqrt{d_1^2 + (\Delta D)^2 -2d_1 |\Delta D| \cos{(\pi/2 + \theta_i)}}& \theta_d \leq 0.
\end{cases}\label{eq:sphere_delay1}
\end{equation}
The delay can be computed as: 
\begin{equation}
\tau_m^{'} = \tau_m + \Delta \tau = \tau_m + (d_2-d_1)/c. \label{eq:sphere_delay2}
\end{equation}

For the plane wave approximation,  the incident angle remains the same when there is a deviation in source position.  However, for spherical wave approximation,  there will be a change in the incident angle as illustrated in Fig.~\ref{Fig:tx_sphere_interp}.   
The new incident angle $\tilde{\theta}_i$ is computed based on  the angle $\gamma_i$ between $d_1$ and $d_2$.  
 The angle $\gamma_i$ is given by:
\begin{equation}
\gamma_i = \arccos{\left(\dfrac{d_1^2 + d_2^2 - |\Delta D|^2}{2d_1 d_2}\right)}.\label{eq:src_interp_gamma}
\end{equation}
Finally, $\tilde{\theta}_i$ is computed as:
\begin{equation}
\tilde{\theta}_i = \theta_i + \mathrm{sign}(\theta_i)\mathrm{sign}(\Delta \tau)\gamma_i.\label{eq:new_theta}
\end{equation}

\subsection{Arbitary position of source and receiver}
Now we consider the general scenario when both the source and the receiver positions deviate from a known depth/grid point. 

The main steps of interpolation are:
\begin{enumerate}
\item Adjust delays of the arrivals at the four grid points based on the deviation of the source depth. For spherical interpolation, use (\ref{eq:sphere_delay1}) and (\ref{eq:sphere_delay2}). For plane wave interpolation,  $\Delta \tau$ is computed using (\ref{eq:plane_src_delay}).
\item For the spherical interpolation only, adjust the incident angles of the arrivals at four grid points based on the deviation of source depth using (\ref{eq:src_interp_gamma}) and (\ref{eq:new_theta}).
\item Adjust delays of the arrivals based on the receiver position between grid points. For spherical interpolation, use (\ref{eq:d3}) - (\ref{eq:gamma}). For plane wave interpolation, use (\ref{eq:plane_rx_delay}).
\end{enumerate}

In subsection~\ref{subsec:receiver_deviate} and \ref{subsec:source_deviate}, we consider specific circumstances when the source/receiver is at a known depth/grid point to breakdown the description of the calculations for arbitary source/receiver positions. In practice, for any source and receiver positions, we follow the interpolation steps summarized in this subsection.

\section{Numerical simulation}\label{sec:numerical_simulation}
In this section, we investigate the performance of the acoustic field interpolation  and show an example of the time-varying CIR when a target crosses a link between a signal source and receiver.

\subsection{Acoustic field interpolation}
In this subsection, the performance of the plane wave and spherical wave acoustic field interpolation methods are compared. 

We consider an environment with a flat surface and a flat bottom. The sea depth is 200~m, and the sound speed profile is the same as in the SWellEx-96 experiment~\cite{booth2000detectability}. Six grid maps are pre-computed for source depths from 15~m to 20~m with a depth step size of $\Delta D_s = 1$~m. The signal source depth in the simulation trials is random and independently uniformly distributed within $[15, 20]$~m. Each grid map covers the receiver range from 1~m to 50~m and receiver depths from 5~m to 40~m. The resolution of the grid maps in both range and depth is 0.5~m. 
In every simulation trial, the receiver range and depth are independent and uniformly distributed within $[15, 20]$~m and $[5, 40]$~m, respectively.

The interpolation accuracy is evaluated by computing the mean squared deviation:
\begin{equation}\nonumber
\mathrm{MSD}_\mathrm{interp} = 10\log_{10}\left(\dfrac{\sum_{n = 0}^{K-1}\left(|\hat{h}(n)| - |h(n)|\right)^2}{\sum_{n = 0}^{K-1}|h(n)|^2}\right),
\end{equation}
where $\mathbf{\hat{h}} = [\hat{h}(0), \ldots, \hat{h}(K-1)]^T$ is a CIR estimate and $\mathbf{h} = [h(0), \ldots, h(K-1)]^T$ is the true CIR. 

The CIR estimate and the true CIR are computed as the inverse FFT of the channel frequency response with the interpolated and true field information, respectively. The interpolated channel frequency response is computed using (\ref{eq:channel_freq_response}) and (\ref{eq:H1}).
For computation of the true CIR, the amplitude and delay of arrivals are generated at the precise positions of the source and receiver by running the BELLHOP3D program. For the CIR estimates, the acoustic field is computed using the plane or spherical wave interpolation as described in Section~\ref{sec:interpolation}.
 In the simulations, the carrier frequency is $f_c = 32$~kHz and the frequency bandwidth is $F_d = 6$~kHz. In total, 10000 simulation trials are run.

\begin{figure}
\centering
 \begin{subfigure}{1\linewidth}
\includegraphics[width=0.9\linewidth]{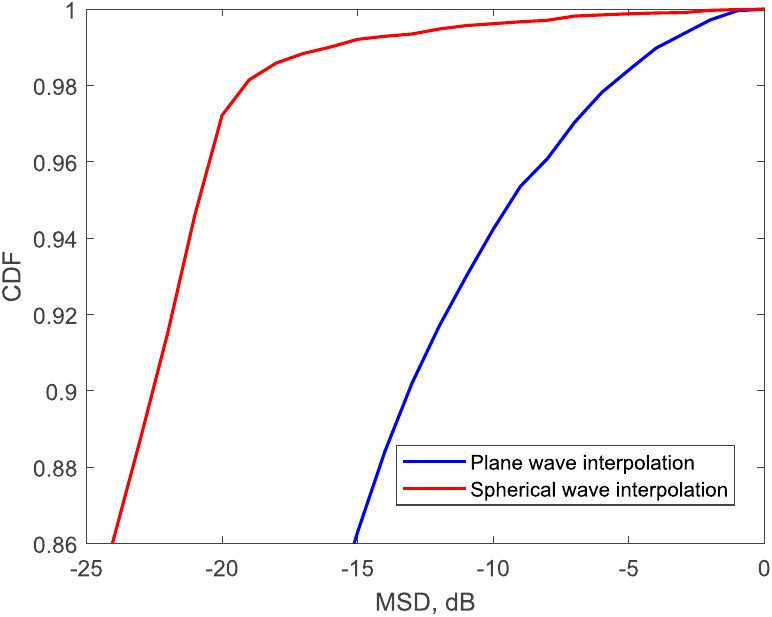}
\caption{\label{Fig:MSD_compare_shortRange}}
\end{subfigure}
\begin{subfigure}{1\linewidth}
\includegraphics[width=0.89\linewidth]{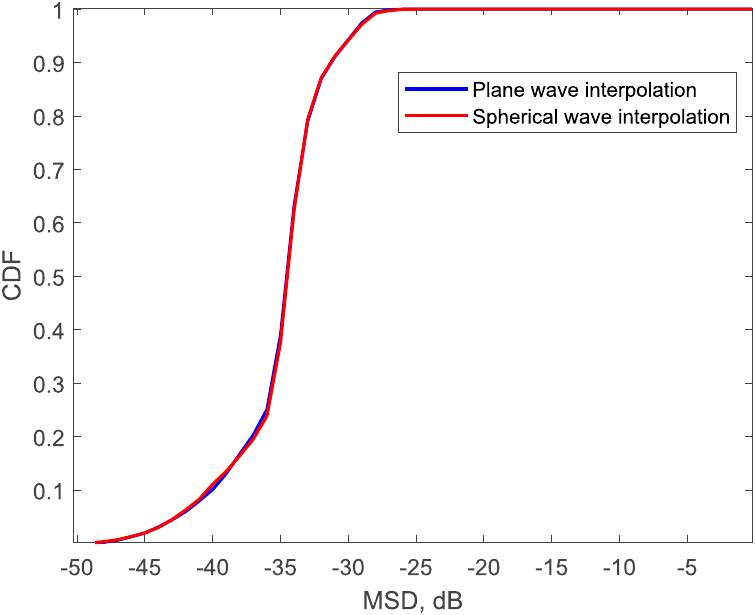}
\caption{\label{Fig:MSD_compare_longRange}}
\end{subfigure}
\caption{CDF of $\mathrm{MSD}_\mathrm{interp}$ for plane wave and spherical wave interpolation: (a) the receiver is at a short range (within 50~m from the signal source); (b) the receiver is at a long range (further than 1000~m away from the signal source).}
\end{figure}

Fig.~\ref{Fig:MSD_compare_shortRange} shows the cumulative distribution function (CDF) of the $\mathrm{MSD}_\mathrm{interp}$ when using the plane wave and spherical wave interpolation. It can be seen that the performance of the spherical wave interpolation is significantly better than that of the plane wave interpolation. An $\mathrm{MSD}_\mathrm{interp}$ better than -15~dB is achieved in $99\%$ of the simulation trials for the spherical interpolation, while only in $86\%$ of the simulation trials for the plane wave interpolation. 

In Fig.~\ref{Fig:MSD_compare_shortRange}, the spherical interpolation shows a significantly better performance than the plane wave interpolation when the receiver is close to the signal source (range smaller than 50~m). 
To investigate the performance at a longer distance, we generate a single grid map at 20~m source depth for a longer range scenario. The signal source depth is randomly chosen within $[19.5, 20.5]$~m. The range of the receiver varies from 1000~m to 1030~m, and its depth varies from 15~m to 25~m. The resolution of the grid map in range and depth are both 1~m. Again, the receiver position is randomly chosen to be within the grid map.  In total, 5000 simulation trials are run. As expected and demonstrated in Fig.~\ref{Fig:MSD_compare_longRange}, almost the same performance is achieved by both plane wave and spherical wave interpolation. 

To conclude, the spherical interpolation is accurate regardless of the range between the source and receiver, which is not the case for the plane wave interpolation. As the trajectory of the target can be close to the source or receiver, the spherical interpolation is a better candidate for the acoustic field interpolation.

\begin{figure}
\centering
\includegraphics[width=0.95\linewidth]{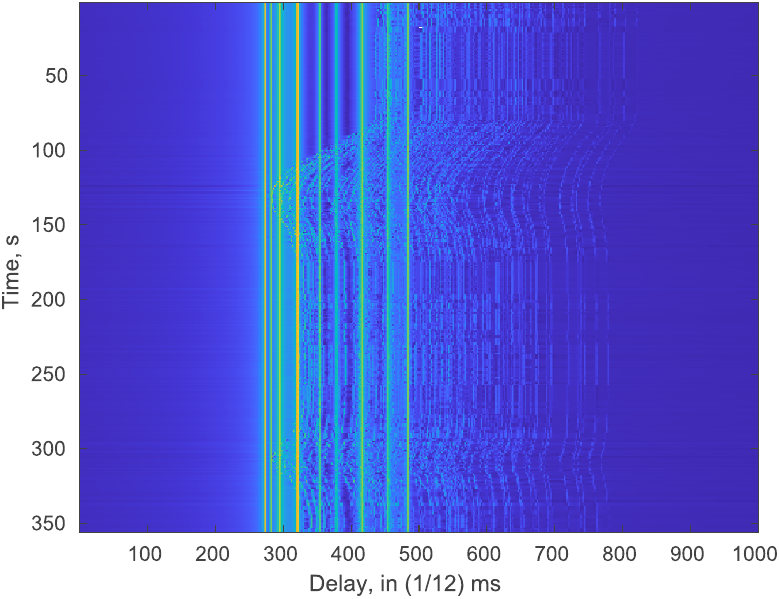}
\caption{Time-varying CIR generated by the simulator based on the acoustic field information. \label{Fig:IR_neptune_sim}}
\end{figure}

\begin{figure}
\centering
\includegraphics[width=0.95\linewidth]{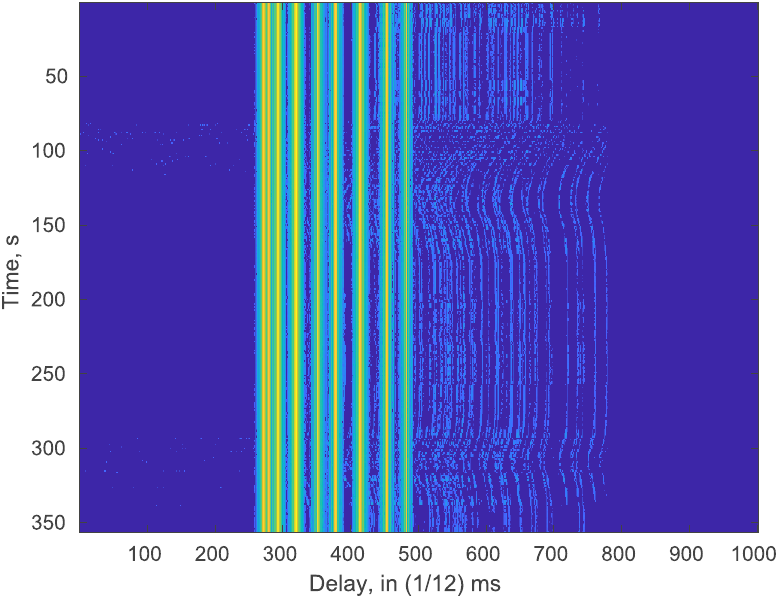}
\caption{Time-varying CIR estimate computed using the received signal. \label{Fig:IR_neptune_computed}}
\end{figure}

\subsection{Time-varying channel impulse response}
In this subsection, we show an example of the time-varying CIR generated by the simulator using the spherical wave interpolation when the target is moving between the signal source and receiver.

This simulation scenario is setup based on one of the communication links in a lake experiment described in Section~\ref{Sec:lake_exp}. For the simulation, we use the first 355~s of the estimated target trajectory shown in Fig.~\ref{Fig:trajectory_est}. The target is moving at a speed of 1~m/s at 1.5~m depth. The distance between two nodes is 14.2~m. The depth of the lake is set as 7~m. The transmitter is at 4.6~m depth and the receiver is at 4.4~m depth. In this simulaton, two grid maps are generated for a source depth of 1.5~m and 4.6~m depth. For each grid map, the range of the receiver varies from 0 to 35~m, and its depth varies from 0 to 7~m. The resolution of the grid map in range and depth is 1~m and 0.5~m, respectively.

As described in Section~\ref{sec:channel_impulse_response}, the CIR is computed as the inverse FFT of the channel frequency response, given by (\ref{eq:channel_freq_response}). The CIR variation is due to the second term in (\ref{eq:channel_freq_response}), where both the links, link 2 and link 3, have time-varying channels. The magnitude of the CIR variation depends on the target strength $\sigma_{TS}$, which is computed using (\ref{eq:target_sphere}). 
For the purpose of observing the additional reflections introduced by the moving target, we set $\sigma_\mathrm{TS}(i) = 1$, i.e., at a high value, to highlight these reflections in the CIR. The time-varying CIR generated by the simulator is shown in Fig.~\ref{Fig:IR_neptune_sim}. A group of static multipath components can be observed between the $250$th and $500$th taps, which represents the acoustic propagation between the source and receiver without the target. A large group of weaker multipath components can be observed with time-varying delays, these are the additional reflections introduced by the moving target. In this simulation scenario, the target crossed the link between the source and the receiver twice. This can be observed from the time-varying delays of the group of reflections, which reduce when the target is approaching the link and increase as the target moving away from the link. 

The time-varying CIR allows us to understand the changes of the CIR  structure when a target is crossing the communication link. 
In practice, channel estimates are obtained by a receiver.
A detailed description of the receiver used in this paper can be found in~\cite{shen2023rake}. One of the features of the receiver, useful for the target detection, is the use of turbo iterations for channel estimation and demodulation, resulting in channel estimates obtained using both the pilot and demodulated symbols, thus significantly improving the estimation accuracy compared to the use of the pilot symbols alone.
The baseband received signal generated by the simulator is used for channel estimation. The time-varying CIR estimate is shown in Fig.~\ref{Fig:IR_neptune_computed}. 
Similar delay variations of the group of refelctions is observed when the target is crossing the communication link.
This demonstrate that the channel estimator in the receiver is capable of tracking the change of the CIR due to a moving target. In Section~\ref{Sec:lake_exp}, this receiver is used for generating the time-varying channel estimate.

The simulator provides a mean to evaluate different algorithms for target detection and has a flexibility of investigating different node configurations within an area of interest. Note that organizing and conducting field experiments with multiple nodes is complicated and expensive; therefore, a simulator for investigation of target detection by a network of nodes is a very important research tool.

With the two-stage processing, the computational time of simulating the acoustic field for one communication link is significantly reduced. 
For this simulation, two grid maps are pre-computed. The computation of each grid map takes 25~minutes. 
For a simulation of 1~s duration, it takes 1.3 minutes to generate the received signal at 12~kHz baseband sampling rate. If the acoustic field is generated directly using BELLHOP3D, it takes about 9 hours to generate 1~s of the received signal with the same sampling rate.



\section{Lake experiment}\label{Sec:lake_exp}
In this section, we describe the lake experiment. Subsection~\ref{sec:exp_config} describes the configuration of the lake experiment. Subsection~\ref{sec:data_packet} introduces the transmitted signal. Subsection~\ref{sec:trajectory} describes the technique used for target trajectory estimation. In subsection~\ref{sec:target_detect}, we demonstrate the target detection capability. Subsection~\ref{sec:exp_sim} demonstrates the target detection performance with the simulated data.

\subsection{Experiment setup}\label{sec:exp_config}

\begin{figure}
      \centering
		\includegraphics[width=1\linewidth]{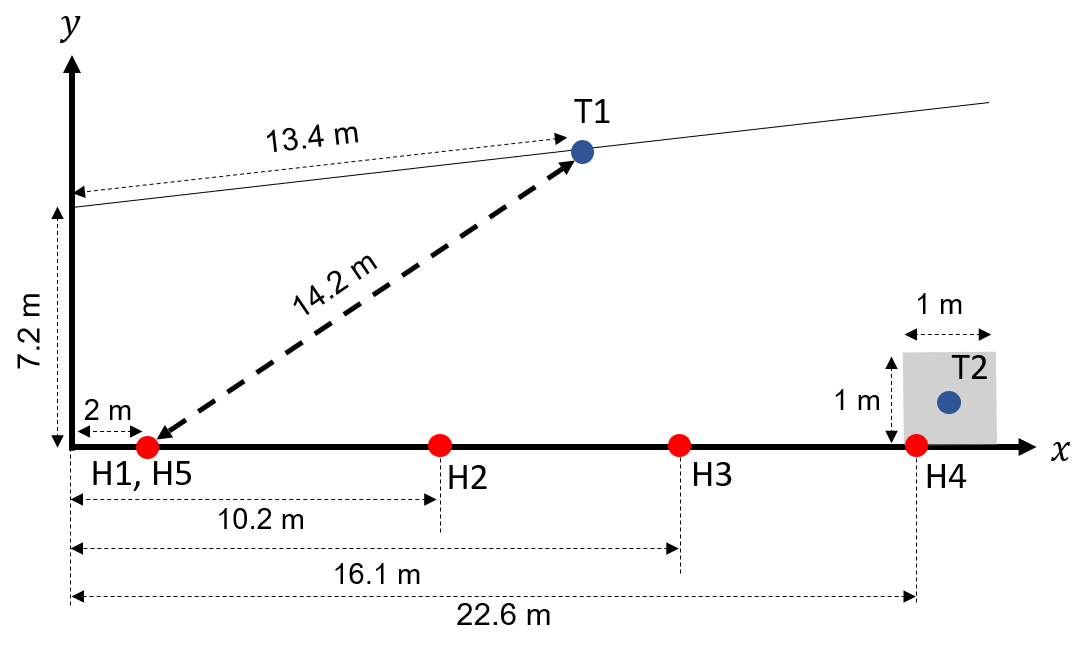}
	\caption{Configuration of the lake experiment.}\label{Fig:exp_config}
\end{figure}

\begin{figure}
      \centering
	     \begin{subfigure}{0.49\linewidth}
		 \includegraphics[trim = {5cm 0 0 0}, clip, width=\linewidth]{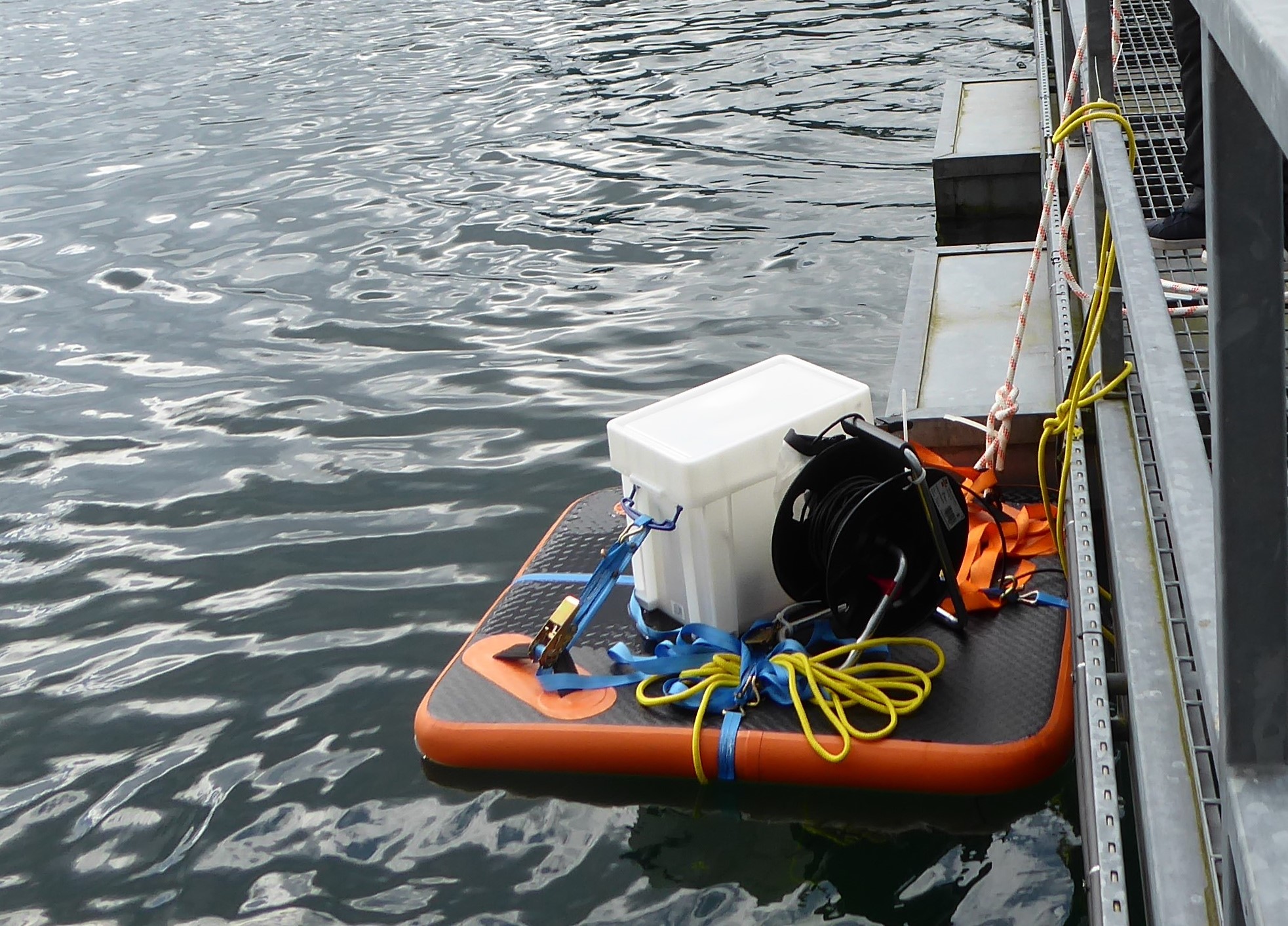}
		 \caption{}
		 \label{fig:platform}
	      \end{subfigure}
	       \begin{subfigure}{0.49\linewidth}
		  \includegraphics[trim = {1.5cm 0 1cm 0}, clip, width=\linewidth]{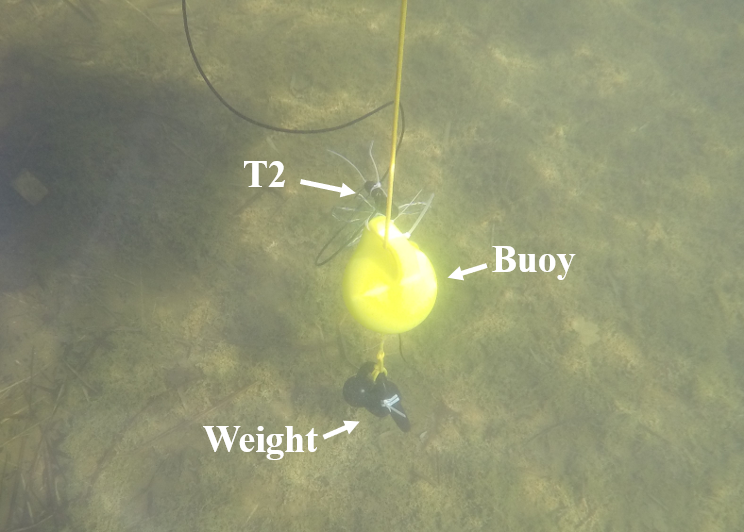}
		  \caption{}
		  \label{fig:target}
	       \end{subfigure}
	\caption{Target deployment: (a) floating platform with the acoustic transmitter operating at the carrier frequency 24~kHz for the target navigation; (b) underwater system attached to the floating platform, including the yellow buoy with a weight and projector T2 connected to the transmitter on the top of the floating platform by a cable.}
	\label{fig:subfigures4}
\end{figure}
The purpose of the lake experiment is to demonstrate the detection of a target crossing a communication link between two nodes.
Fig.~\ref{Fig:exp_config} shows the configuration of the lake experiment. 
Two projectors (T1 and T2) and five hydrophones (H1 to H5) are deployed.
T1 is the source node transmitting data packets of 0.1~s duration and 1~s period at $32$~kHz carrier frequency, while T2 is fixed on the target (the yellow buoy in Fig.~\ref{fig:target}) transmitting data packets at $24$~kHz carrier frequency for the purpose of the acoustic navigation. T1 is fixed at 4.6~m depth. 
Hydrophones H2 to H5 are fixed at a depth of 1.5~m above the lake bottom. Hydrophone H1 is fixed at a depth of 3~m above the lake bottom. The depth of the lake site varies from 5~m to 8~m depth.

The target is fixed underneath a $1\mathrm{m} \times 1\mathrm{m}$ floating platform shown in Fig.~\ref{fig:platform}. A weight is attached to the target to ensure it is at approximately 1.5~m depth. During the experiment, a rope is used to pull the floating platform with the target attached as shown in Fig.~\ref{fig:target} diagonally from the point $(22.7, 0)$, near the hydrophone H4, to the point $(1, 7.2)$, or in the opposite direction. The speed of the target movement is about 1~m/s. This target movement was repeated several times during the experiment.

\subsection{Data packet transmission}\label{sec:data_packet}
We consider data packet transmission in the network.
A superimposed pilot and data packet structure is used, with pilot
symbols $p(i)$ in the real part and data symbols $d(i)$ in the imaginary part~\cite{shen2023rake}. Both the pilot and data are binary phase-shift keying (BPSK) symbols. The data symbols are encoded by a rate 1/3 convolutional code and interleaved. For the transmitted signal, single-carrier modulation is considered. The superimposed pilot and data symbols are pulse-shape filtered and up-sampled by a root raised-cosine (RRC) filter with a roll-off factor of 0.2. After pulse-shaping, the signal is up-converted to the carrier frequency. 

\subsection{Target trajectory estimation}\label{sec:trajectory}
To evaluate the target detection performance, the ground truth of the target trajectory is required. 
To accurately estimate the trajectory of the target during the experiment, the projector T2 attached to the target transmits the target navigation signal in a frequency band outside of that of the communication signals for the target detection (see Fig.~\ref{fig:target}). A simple navigation algorithm is then used based on estimating delays of the first arrivals at the multiple receiving nodes H1 to H5. With the signal bandwidth $F_d = 6$~kHz and oversampling at the receiver to 12~kHz, the delay resolution is 1/12~ms, thus the distance resolution is about 0.1~m, which is good enough for the purpose of the experiment.

\begin{figure}
\centering
\begin{subfigure}{1.01\linewidth}
    \includegraphics[width=1\linewidth]{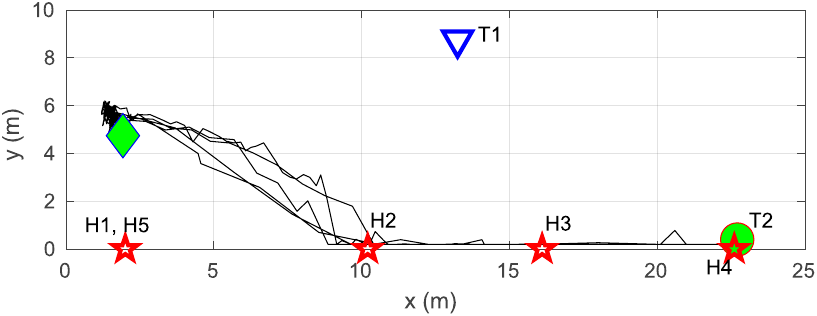}
    \caption{}\label{fig:traj_est_a}
\end{subfigure}
\begin{subfigure}{1\linewidth}
\vspace{5pt}
    \includegraphics[width=1\linewidth]{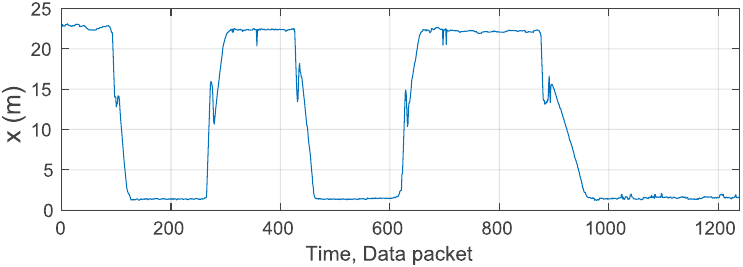}
    \caption{}\label{fig:traj_est_b}
\end{subfigure}
\hfill
\begin{subfigure}{1\linewidth}
\vspace{5pt}
    \includegraphics[width=1\linewidth]{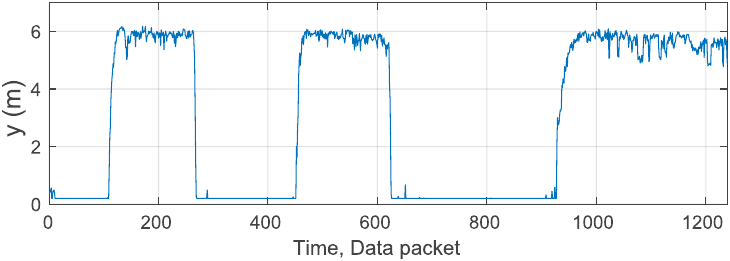}
    \caption{}\label{fig:traj_est_c}
\end{subfigure}
\caption{Coordinate estimates of the target; (a) trajectory of the target during the experiment; (b) $x$ coordinate estimates; (c) $y$ coordinate estimates. \label{Fig:trajectory_est}}
\end{figure}

Fig.~\ref{fig:traj_est_a} shows the estimated trajectory of the target during the experiment as a black curve. The green circle shows the starting point and the green diamond shows the final position of the target.  Fig.~\ref{fig:traj_est_b}  and \ref{fig:traj_est_c} show how the target coordinates $x$ and $y$, respectively, vary in time. Each slope in Fig.~\ref{fig:traj_est_b} and ~\ref{fig:traj_est_c} corresponds to a crossing of the target between the links T1 and H1/H5. The other links are not useful, since the target moves very closely to the other receivers (H2 to H4). Therefore, we use only these two communication links for target detection.

\subsection{Target detection}\label{sec:target_detect}

\begin{figure}
\centering
\includegraphics[width=1\linewidth]{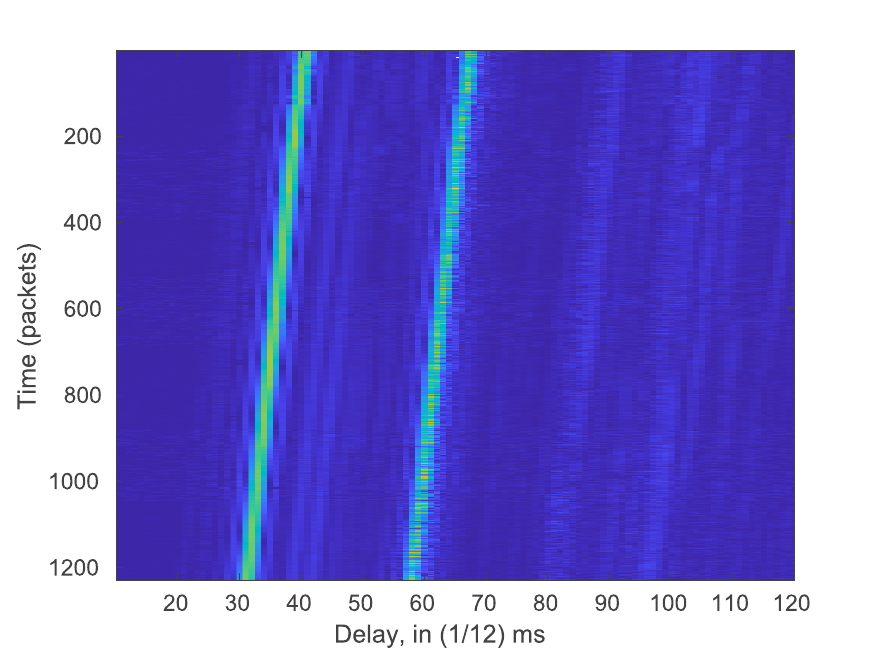}
\caption{Channel impulse response between T1 and H5. \label{Fig:exp_IR_T2R5}}
\end{figure}

The first step of target detection is to obtain the CIR estimates between T1 and H1 and between T1 and H5, these are available at the receivers. 
As an example, we show the estimates of the CIR between T1 and H5 in Fig.~\ref{Fig:exp_IR_T2R5}. It can be observed that there is a shift in the delay of the multipath structure over time. This is caused by the unsynchronized clocks on recorders used for data transmission and reception. During the whole transmission interval of about 20 minutes, the delay of the direct path shifted by 0.8~ms. From Fig.~\ref{Fig:exp_IR_T2R5}, it is not possible to visually identify the CIR changes corresponding to the target crossing the communication link.

\begin{figure*}
\centering
\begin{subfigure}{0.78\linewidth}
    \includegraphics[trim = {0cm 0.1cm 0cm 0.1cm}, clip, width=1\linewidth]{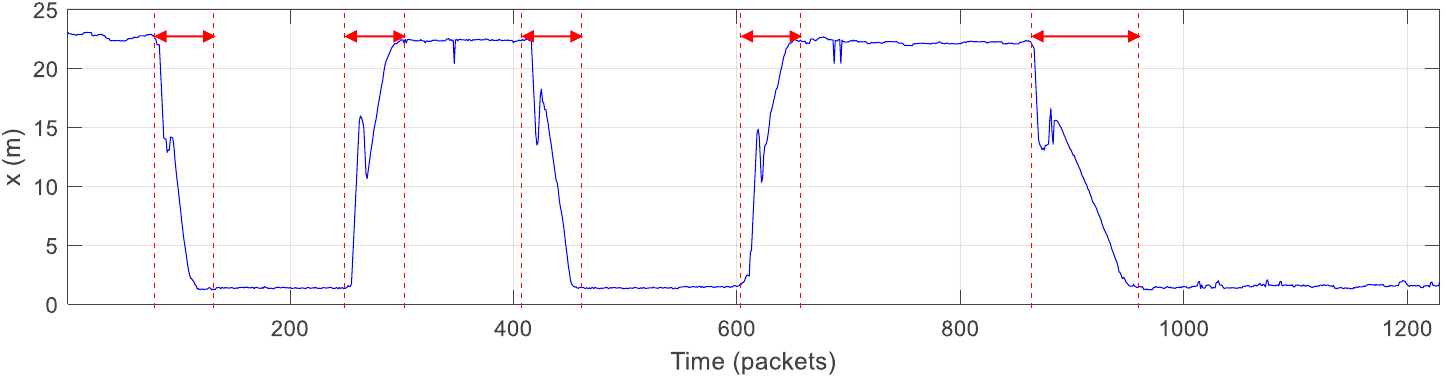}
    \caption{}
\end{subfigure}
\begin{subfigure}{0.79\linewidth}
    \includegraphics[trim = {2.6cm 0.1cm 2.6cm 0.2cm}, clip, width=1\linewidth]{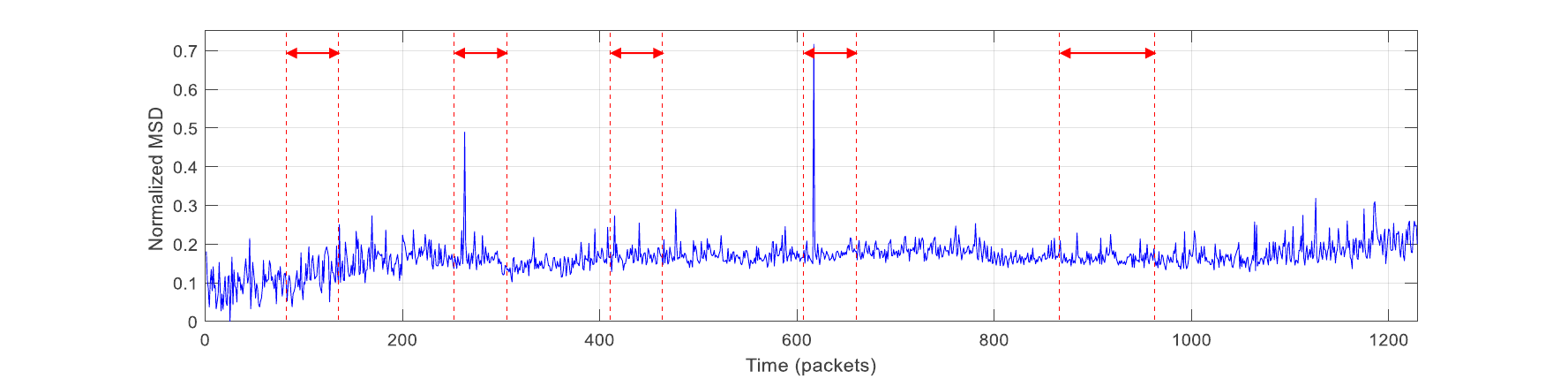}
    \caption{\label{fig:ce_rx1}}
\end{subfigure}
\begin{subfigure}{0.79\linewidth}
    \includegraphics[trim = {2.6cm 0.1cm 2.6cm 0.2cm}, clip, width=1\linewidth]{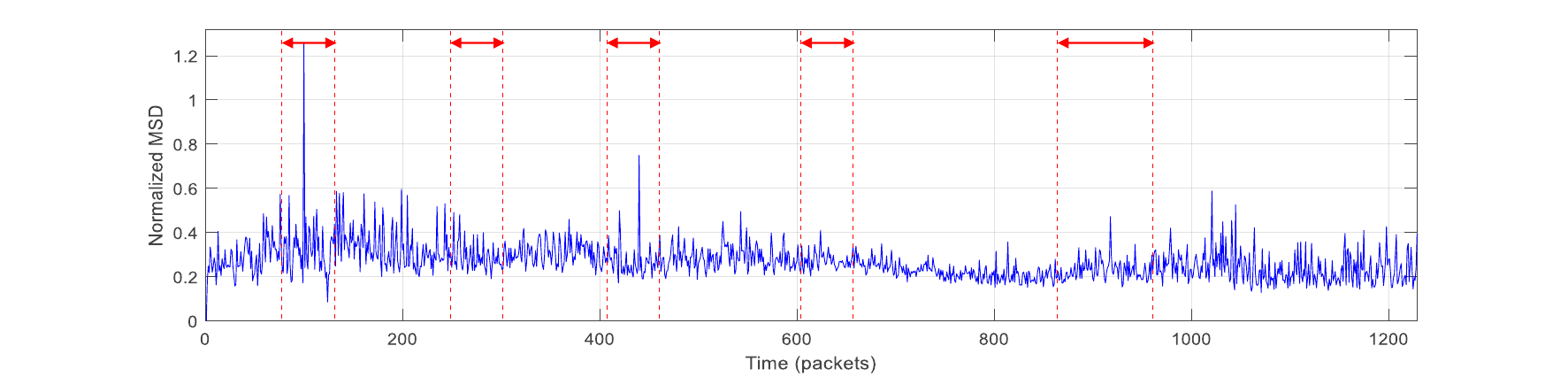}
    \caption{\label{fig:ce_rx5}}
\end{subfigure}
\caption{Normalized MSD in the experiment; (a) $x$-coordinate estimates; (b) link between T1 and H1; (c) link between T1 and H5. \label{Fig:ce_exp}}
\end{figure*}
\begin{figure*}
\centering
\begin{subfigure}{0.75\linewidth}
\centering
\includegraphics[trim = {2.45cm 0.1cm 2.7cm 0.1cm}, clip,width=1\linewidth]{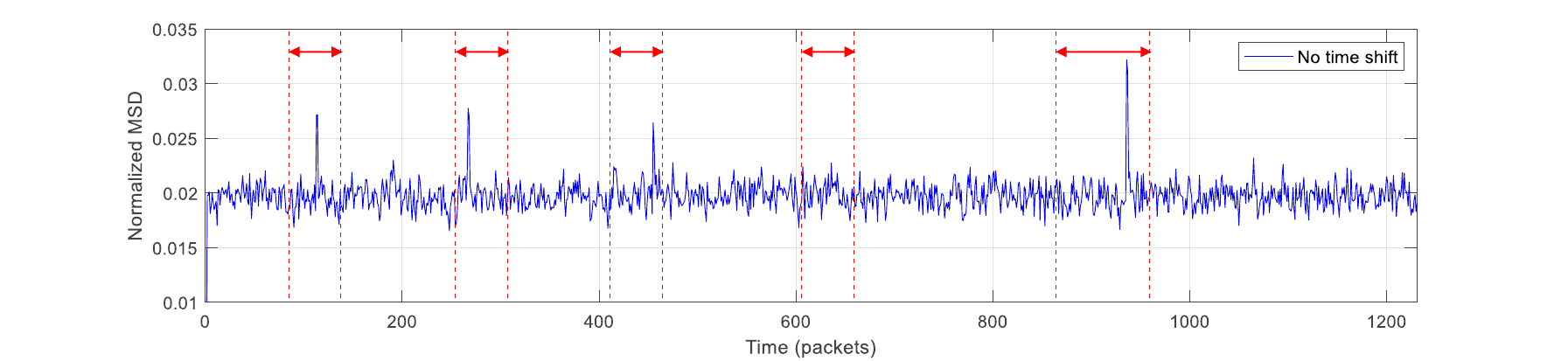}
\caption{\label{Fig:exp_sim1}}
\end{subfigure}
\begin{subfigure}{0.75\linewidth}
\centering
\includegraphics[trim = {2.45cm 0.1cm 2.7cm 0.1cm}, clip, width=1\linewidth]{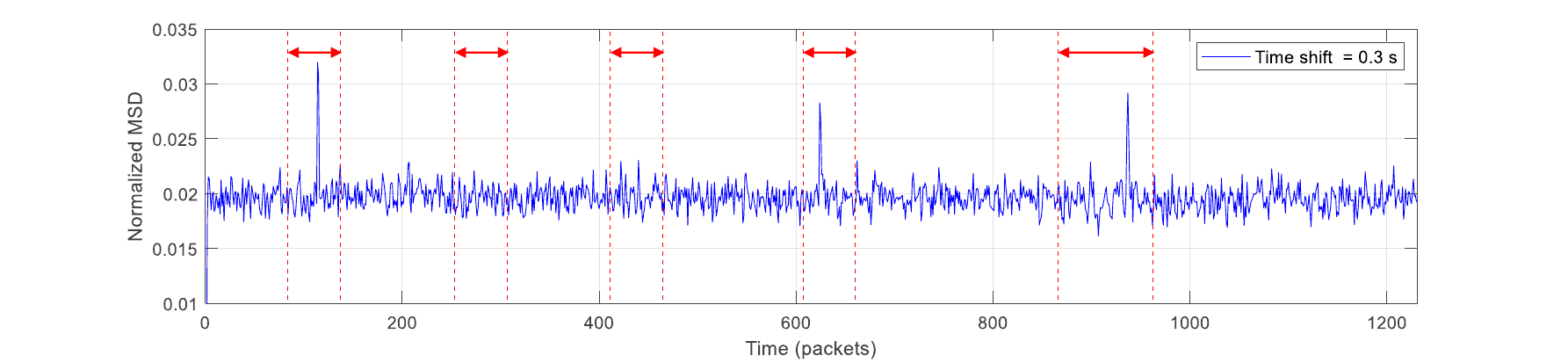}
\caption{ \label{Fig:exp_sim2}}
\end{subfigure}
\caption{Normalized MSD with the simulated data: (a) Data transmission starts from zero time instant; (b) Data transmission starts from 0.3 s.\label{Fig:exp_sim}}
\end{figure*}
The normalized MSD during the experiment for all 1200 transmitted data packets according to (\ref{eq:norm_MSD}) is plotted in Fig.~\ref{fig:ce_rx1} and \ref{fig:ce_rx5} for the receivers H1 and H5, respectively. We use the first CIR  in the experiment, when the target was in the position $(22.7, 0)$, far from the communication links of interest, as the reference impulse response. The parameters used in the filter $g$ in (\ref{eq:G_k}) are: $P = 15$, $K = 2048$. The regularization parameter is $\varepsilon = 10^{-3}$. A large number of taps in filter $g$ is used to compensate for the delay shift due to the unsynchronized clocks in the recorders and environment variation.

The potential time window of the target crossing is marked with red dotted lines and arrows in Fig.~\ref{Fig:ce_exp}. These time windows are chosen based on the changes in the $x$ coordinate estimates. For the channel between T1 and H1, two crossings are detected at the 263th and 617th data packets. For the channel between T1 and H5, two crossings are detected at the 100th and 440th data packets. Due to the one second gap in the data packet transmission, it is possible to miss the detection of the target crossing. In this experiment, four out of five target crossings are clearly detected based on the CIRs from H1 and H5. This could be further improved by deploying more hydrophones at different depths and/or reducing the gap between data packet transmissions.

\subsection{Experiment-based simulation}\label{sec:exp_sim}
This subsection shows the target detection performance obtained with the simulated data generated based on the setup in lake experiment.

We simulate the channel between T1 and H1 for demonstration. During the experiment, T1 and H1 are located at 4.6~m and 5.9~m depth, respectively. We assume that the target is moving at a depth of 1.5~m. The coordinate estimates shown in Fig.~\ref{Fig:trajectory_est} are used for interpolating the trajectory of the target during the experiment. The depth of the lake is set to 7~m. Two grid maps are generated at source depths of 1.5~m and 4.6~m. Each grid map covers an area of $35~\mathrm{m}$ in range and $7~\mathrm{m}$ in depth. The resolution of the grid map is 1~m in range and 0.5~m in depth. The sound speed in the lake is set based on the temperature of the water ($10^{\circ}$) as $c = 1443$~m/s.

The target strength $\sigma_{TS}$ is computed using (\ref{eq:target_sphere}) with the radius of the target set to 0.1~m. Random Gaussian noise is added to the time-varying CIRs to limit the CIR estimation accuracy at a level of -20~dB in terms of MSD. As there is no issue with the unsynchronized clock in the simulation, we use $P = 1$. 

Fig.~\ref{Fig:exp_sim} shows the normalized MSD in two experiments. In the first experiment, the data packets are transmitted from zero time instant, while for the second experiment, the transmission starts from 0.3~s. It can be seen that four of five target crossings are detected in the first experiment. In the second experiment three crossings are detected. This is similar to what is observed in the lake experiment. The simulation results demonstrate that the target detection performance is sensitive to the timing of data packet transmissions with respect to the target crossing. To increase the possibility of target detection, more frequent data packet transmission is required.

\section{Conclusions}\label{sec:conclude}
In this paper, we have investigated the possibility of target detection by using data transmission between communication nodes in an UWA network.
A target detection method has been proposed based on the changes in the time-varying CIRs with respect to a reference CIR obtained without a target. An acoustic simulator has been developed, which is capable of virtual transmission of acoustic signals through a communication link in the presence of a moving target and an approach has been proposed to improve the acoustic field interpolation in the near field. The performance of the target detection method has been investigated using numerical simulation and lake experiments. 
The proposed detection metric allows clear indication of a moving target crossing a communication link. 

\bibliographystyle{IEEEtran}
\bibliography{exp_sim_target_detection}

\end{document}